\documentclass[a4paper,aps,twocolumn,nofootinbib]{revtex4}
\RequirePackage[colorlinks,hyperindex]{hyperref}
\RequirePackage[english]{babel}
\RequirePackage[latin1]{inputenc}
\RequirePackage[T1]{fontenc}
\RequirePackage{mathrsfs}
\RequirePackage{amsmath}
\RequirePackage{amssymb}
\RequirePackage{amsbsy}
\RequirePackage{color}
\RequirePackage{bm}
\hypersetup{colorlinks=true,breaklinks=true,urlcolor=blue,linkcolor=blue}
\begin{document}
\title{\bf{Torsion Axial Vector and Yvon-Takabayashi Angle:\\ 
Zitterbewegung, Chirality and all that}}
\author{Luca Fabbri$^{\ast}$, Rold\~{a}o da Rocha$^{\dagger}$}
\affiliation{$^{\ast}$ DIME, Universit\`{a} di Genova,\\ P.Kennedy Pad.D, 16129 Genova, ITALY,\\
$^{\dagger}$ CMCC, Universidade Federal do ABC,\\ 09210-580, Santo Andr\'e, BRAZIL}
\date{\today}
\begin{abstract}
We consider propagating torsion as a completion of gravitation in order to describe the dynamics of curved-twisted space-times filled with Dirac spinorial fields; we discuss interesting relationships of the torsion axial vector and the curvature tensor with the Yvon-Takabayashi angle and the module of the spinor field, that is the two degrees of freedom of the spinor field itself: in particular, we shall discuss in what way the torsion axial vector could be seen as the potential of a specific interaction of the Yvon-Takabayashi angle, and therefore as a force between the two chiral projections of the spinor field itself. Chiral interactions of the components of a spinor may render effects of zitterbewegung, as well as effective mass terms and other related features: we shall briefly sketch some of the analogies and differences with the similar but not identical situation given by the Yukawa interaction occurring in the Higgs sector of the standard model. We will provide some overall considerations about general consequences for contemporary physics, consequences that have never been discussed before, so far as we are aware, in the present physics literature.
\end{abstract}
\maketitle
\section{Introduction}
The Dirac spinorial field theory has the spinor field as fundamental ingredient, defined in terms of its spinorial transformation law: the classical Dirac spinor is a column of $4$ complex scalar fields, carrying spin-$\frac{1}{2}$ representations of the Lorentz group; this means that the Dirac spinor has in total $8$ real components, of which $6$ removable by a Lorentz transformation, and so that it has $2$ real degrees of freedom alone \cite{L,Cavalcanti:2014wia,Fabbri:2016msm}. Essentially, the idea is that the $6$ removable components are the $3$ components of the space velocity, removable by boosts, and the $3$ components of the spin, removable by rotations: the two remaining components are the physical degrees of freedom and they are the module and what in literature is known as the Yvon-Takabayashi angle \cite{j-l,h1,Fabbri:2017pwp}. On the other hand, however, in any treatment of modern physics, like QFT for instance, after having boosted into the rest frame and after having rotated the spinor so to have it aligned along the third axis, the final form of the spinor field is a constant.

This apparent discrepancy comes from the fact that in modern treatments all we use are plane waves, for which the module is constant and the YT angle is zero \cite{p-s}; still, we know that in the most general of cases there must be a varying module and a non-zero YT angle. So in general we know that a given quantity is present, although in the commonly accepted paradigms it is always set to zero.

A similar situation occurs when the Dirac theory is endowed with a dynamics, that is the Dirac spinor field is seen as solution of the Dirac equation and with conserved quantities as source of geometric field equations: a Dirac spinor field possesses three conserved quantities, that is the current and energy density as well as the spin density.

On the other hand, the geometric field equations come from the set-up of a geometrical background, and in general it is possible to see electrodynamics emerging as a gauge theory of the unitary phase transformation similarly to the fact that gravity emerges from the non-trivial structure of the space-time: when the background is constructed in its full generality, we witness the appearance of gauge strengths beside curvature as well as torsion \cite{Fabbri:2017fac}.

Like in Maxwell theory one takes the charge as source of electrodynamic field equations, analogously in Einstein gravitation one takes the energy as source of gravitational field equations and in the same spirit one should take the spin as source of torsion field equations. But again in the usual treatment, as in the case of QFT, this is never done.

An apparent discrepancy emerges once again; and once again we may claim that in the most general case the spin exists and should be coupled, and it is natural to have it coupled to the torsion of the space-time. This is a second instance in which in general some dynamics is to be given although in common paradigms it is always absent.

As we will have the opportunity to discuss later on, the YT angle will indeed be connected to the spin of a spinor field, and henceforth there will be relationships between the YT angle and the torsion. This connection might well justify why in the normal paradigms neglecting both can be equivalent to neglecting only one of these two.

Still, even neglecting one of them is an artificial constraint, and it is interesting to study what happens when, in the most general circumstance, both torsion and YT angle are allowed to take place in the dynamics.

This is what we will do next in this paper.
\section{Geometry of the Spinor Fields}
We begin by recalling the notation we use: throughout the paper, the metric tensor $g_{\alpha\rho}$ defines the orthonormal tetrads $e^{\alpha}_{a}$ so that $g_{\alpha\rho}e^{\alpha}_{a}e^{\rho}_{b}\!=\!\eta_{ab}$ with $\eta_{ab}$ a Minkowskian matrix; matrices $\boldsymbol{\gamma}^{a}$ belong to the Clifford algebra so that the relationships $\left[\boldsymbol{\gamma}^{a}\!,\!\boldsymbol{\gamma}^{b}\right]\!=\! 4\boldsymbol{\sigma}^{ab}$ and $2i\boldsymbol{\sigma}_{ab}\!=\!\varepsilon_{abcd}\boldsymbol{\pi}\boldsymbol{\sigma}^{cd}$ define $\boldsymbol{\sigma}^{ab}$ matrices and the $\boldsymbol{\pi}$ matrix\footnote{This matrix is what is usually indicated as gamma with an index five, but since in the space-time this index has no meaning we prefer to use a notation with no index at all.} implicitly. We use the metric to raise/lower Greek indices, tetrads to change Greek into Latin indices and the Minkowskian matrix to raise/lower Latin indices; procedure $\overline{\psi}\!=\!\psi^{\dagger}\boldsymbol{\gamma}_{0}$ is used to pass from spinors to conjugate spinors in such a way that
\begin{eqnarray}
&2i\overline{\psi}\boldsymbol{\sigma}^{ab}\psi\!=\!M^{ab}\label{1}\\
&\overline{\psi}\boldsymbol{\gamma}^{a}\boldsymbol{\pi}\psi\!=\!S^{a}\label{2}\\
&\overline{\psi}\boldsymbol{\gamma}^{a}\psi\!=\!U^{a}\label{3}\\
&i\overline{\psi}\boldsymbol{\pi}\psi\!=\!\Theta\label{4}\\
&\overline{\psi}\psi\!=\!\Phi\label{5}
\end{eqnarray}
are all real tensor quantities, and they verify the identities
\begin{eqnarray}
&M_{ab}(\Phi^{2}\!+\!\Theta^{2})\!=\!\Phi U^{j}S^{k}\varepsilon_{jkab}\!+\!\Theta U_{[a}S_{b]}
\end{eqnarray}
and
\begin{eqnarray}
&U_{a}U^{a}\!=\!-S_{a}S^{a}\!=\!|\Theta|^{2}\!+\!|\Phi|^{2}\label{norm}\\
&U_{a}S^{a}=0\label{ort}
\end{eqnarray}
called Fierz identities. The celebrated Lounesto classification splits classical spinor fields in terms of the above bi-linear quantities into six altogether exhaustive but mutually exclusive classes \cite{L}, the first three containing the regular spinor fields, for which either $\Theta$ or $\Phi$ are not zero, the second three containing the singular spinor fields, for which both $\Theta$ and $\Phi$ are equal to zero identically.

In the special case in which $\Theta\!=\!\Phi\!=\!0$ spinor fields are called singular and they constitute a class of very interesting spinor fields \cite{daSilva:2012wp, Ablamowicz:2014rpa, daRocha:2016bil, daRocha:2008we, Villalobos:2015xca, Cavalcanti:2014uta,daRocha:2013qhu}. Here however we will focus on the most general case where either $\Theta$ or $\Phi$ are non-zero: in this case (\ref{norm}) tells us that $U^{a}$ is time-like, so that it is always possible to perform three boosts to bring its space part to vanish; then (\ref{ort}) tells us that $S^{a}$ has only its space part, which can always be aligned with the third axis by employing two rotations. It is easy to see that the most general spinor compatible with these constraints is either
\begin{eqnarray}
&\!\psi\!=\!\left(\!\begin{tabular}{c}
$\pm e^{\frac{i}{2}\beta}$\\
$0$\\
$e^{-\frac{i}{2}\beta}$\\
$0$
\end{tabular}\!\right)\!e^{-i\alpha}\phi\ \ \ \ \mathrm{or}
\ \ \ \ \psi\!=\!\left(\!\begin{tabular}{c}
$0$\\
$\pm e^{\frac{i}{2}\beta}$\\
$0$\\
$e^{-\frac{i}{2}\beta}$
\end{tabular}\!\right)\!e^{-i\alpha}\phi
\end{eqnarray}
according to whether the axial vector is either aligned or anti-aligned with the third axis and where the sign ambiguity corresponds to the fact that either $\psi$ or $\boldsymbol{\pi}\psi$ can be considered; this expression has been obtained performing only spinor transformations $\boldsymbol{S}$ and so $\boldsymbol{S}^{-1}\psi$ is the most general form of spinor. Writing $\boldsymbol{S}$ explicitly, it is a simple exercise to see that the most general spinor is therefore
\begin{eqnarray}
&\!\psi\!=\!\phi\sqrt{\frac{2}{\gamma+1}}e^{-i\alpha}e^{-\frac{i}{2}\beta\boldsymbol{\pi}}
\!\left(\!\begin{tabular}{c}
$\left(\frac{\gamma+1}{2}\boldsymbol{\mathbb{I}}\!-\!
\gamma\vec{v}\!\cdot\!\vec{\frac{\boldsymbol{\sigma}}{2}}\right)\xi$\\
$\left(\frac{\gamma+1}{2}\boldsymbol{\mathbb{I}}\!+\!
\gamma\vec{v}\!\cdot\!\vec{\frac{\boldsymbol{\sigma}}{2}}\right)\xi$
\end{tabular}\!\right)
\label{spinor}
\end{eqnarray}
up to the change of $\psi$ into $\boldsymbol{\pi}\psi$ and in which we have that $\gamma\!=\!1/\!\sqrt{1\!-\!v^{2}}$ is the relativistic factor\footnote{This factor should not be confused with the Clifford matrices.} given in terms of the velocity $\vec{v}$ while $\xi$ such that $\xi^{\dagger}\xi\!=\!1$ is some arbitrary semi-spinor field and $\alpha$ is a generic unitary phase.

When either $\Theta$ or $\Phi$ are non-zero we can invert
\begin{eqnarray}
&M_{ab}\!=\!(\Phi^{2}\!+\!\Theta^{2})^{-1}(U^{j}S^{k}\varepsilon_{jkab}\Phi\!+\!U_{[a}S_{b]}\Theta)
\label{M}
\end{eqnarray}
and by defining the directions
\begin{eqnarray}
&S^{a}\!=\!(\Phi^{2}\!+\!\Theta^{2})^{\frac{1}{2}}s^{a}\label{S}\\
&U^{a}\!=\!(\Phi^{2}\!+\!\Theta^{2})^{\frac{1}{2}}u^{a}\label{U}
\end{eqnarray}
we have that
\begin{eqnarray}
&s^{a}\!=\!\frac{1}{\gamma+1}\left(\!\begin{tabular}{c}
$2\gamma(\gamma\!+\!1)(\vec{v}\!\cdot\!\vec{\varsigma})$\\
$2(\gamma+1)\vec{\varsigma}\!+\!2\gamma^{2}(\vec{v}\!\cdot\!\vec{\varsigma})\vec{v}$
\end{tabular}\!\right)\label{s}\\
&u^{a}\!=\!\frac{1}{\gamma+1}\left(\!\begin{tabular}{c}
$\frac{1}{2}(\gamma\!+\!1)^{2}+\frac{1}{2}\gamma^{2}(\vec{v}\!\cdot\!\vec{v})$\\
$\gamma(\gamma\!+\!1)\vec{v}$
\end{tabular}\!\right)\label{u}
\end{eqnarray}
where $\xi^{\dagger}\vec{\boldsymbol{\sigma}}\xi\!=\!2\vec{\varsigma}$ is the spin and such that (\ref{norm}, \ref{ort}) become $u_{a}u^{a}\!=\!-s_{a}s^{a}\!=\!1$ and $u_{a}s^{a}\!=\!0$ with 
\begin{eqnarray}
&\Theta\!=\!2\phi^{2}\sin{\beta}\label{T}\\
&\Phi\!=\!2\phi^{2}\cos{\beta}\label{F}
\end{eqnarray}
in terms of the scalar and pseudo-scalar $\phi$ and $\beta$ that we have defined above: $u^{a}$ is the velocity and $s^{a}$ is the spin, while $\phi$ represents the module and $\beta$ represents the phase difference between the two chiral parts of a spinor field and it is known in literature as Yvon-Takabayashi angle.

From the metric we define $\Lambda^{\sigma}_{\alpha\nu}$ as the symmetric connection, and with it $\Omega^{a}_{b\pi}\!=\!\xi^{\nu}_{b}\xi^{a}_{\sigma}(\Lambda^{\sigma}_{\nu\pi}\!-\!\xi^{\sigma}_{i}\partial_{\pi}\xi_{\nu}^{i})$ is the spin connection in general; then it is possible to define
\begin{eqnarray}
&\boldsymbol{\Omega}_{\mu}
=\frac{1}{2}\Omega^{ab}_{\phantom{ab}\mu}\boldsymbol{\sigma}_{ab}
\!+\!iqA_{\mu}\boldsymbol{\mathbb{I}}\label{spinorialconnection}
\end{eqnarray}
in terms of the spin connection and the gauge potential of charge $q$ and called spinorial connection. Remark that for now we have only defined the torsionless connection.

With the spinorial covariant derivative of (\ref{spinor}) we get
\begin{eqnarray}
\nonumber
&\boldsymbol{\nabla}_{\mu}\psi\!=\![\nabla_{\mu}\ln{\phi}\mathbb{I}
\!-\!\frac{i}{2}\nabla_{\mu}\beta\boldsymbol{\pi}+\\
&+i(qA_{\mu}\!-\!\nabla_{\mu}\alpha)\mathbb{I}
\!+\!\frac{1}{2}(\Omega_{ij\mu}\!-\!\partial_{\mu}\theta_{ij})\boldsymbol{\sigma}^{ij}]\psi
\label{decspinder}
\end{eqnarray}
where $\alpha$ is the unitary phase and the $\theta_{ij}$ encode the three rapidities expressed through the velocity $\vec{v}$ as well as the three rotations expressed as coefficients of the semi-spinor $\xi$ such that $\xi^{\dagger}\xi\!=\!1$ in general: from this we also have 
\begin{eqnarray}
&\nabla_{\mu}s_{\alpha}\!=\!(\partial\theta\!-\!\Omega)_{\rho \alpha\mu}s^{\rho}\label{ds}\\
&\nabla_{\mu}u_{\alpha}\!=\!(\partial\theta\!-\!\Omega)_{\rho\alpha\mu}u^{\rho}\label{du}
\end{eqnarray}
from which we can also calculate all the divergences and curls of these vectors in every equation that follows.

For the dynamics, we assume the action given by
\begin{eqnarray}
\nonumber
&\mathscr{L}\!=\!\frac{1}{4}(\partial W)^{2}\!-\!\frac{1}{2}M^{2}W^{2}
\!+\!R\!+\!\frac{1}{4}F^{2}-\\
&-i\overline{\psi}\boldsymbol{\gamma}^{\mu}\boldsymbol{\nabla}_{\mu}\psi
\!+\!XS^{\mu}W_{\mu}\!+\!m\Phi
\label{l}
\end{eqnarray}
in which $(\partial W)_{\mu\nu}$ is the curl of $W_{\mu}$ being the torsion axial vector with $R$ Ricci scalar and $F_{\mu\nu}$ Faraday tensor, and where $X$ is the strength of the interaction between torsion and the spin of spinor fields while $M$ and $m$ are the mass of torsion and the spinor field itself. Having defined the connection in the torsionless case it would seem we went in a case of restricted generality, but in reality we are still in the most general situation even if the connection has no torsion so long as torsion is eventually included in the form of a supplementary massive axial vector field \cite{Fabbri:2014dxa}.

Varying the above Lagrangian functional with respect to the spinor field gives the Dirac spinor field equations according to the usual form supplemented by a torsional interaction with the spinor field itself according to
\begin{eqnarray}
&i\boldsymbol{\gamma}^{\mu}\boldsymbol{\nabla}_{\mu}\psi
\!-\!XW_{\mu}\boldsymbol{\gamma}^{\mu}\boldsymbol{\pi}\psi\!-\!m\psi\!=\!0\label{D}
\end{eqnarray}
as field equations for matter. Dotting by $\boldsymbol{\gamma}^{a}\boldsymbol{\pi}$ and $\boldsymbol{\gamma}^{a}$ and by the conjugate spinor, and splitting real and imaginary parts, gives the four tensor real decompositions
\begin{eqnarray}
\nonumber
&i(\overline{\psi}\boldsymbol{\nabla}^{\alpha}\psi
\!-\!\boldsymbol{\nabla}^{\alpha}\overline{\psi}\psi)
\!-\!\nabla_{\mu}M^{\mu\alpha}-\\
&-XW_{\sigma}M_{\mu\nu}\varepsilon^{\mu\nu\sigma\alpha}\!-\!2mU^{\alpha}\!=\!0
\label{vr}\\
\nonumber
&\nabla_{\alpha}\Phi
\!-\!2(\overline{\psi}\boldsymbol{\sigma}_{\mu\alpha}\!\boldsymbol{\nabla}^{\mu}\psi
\!-\!\!\boldsymbol{\nabla}^{\mu}\overline{\psi}\boldsymbol{\sigma}_{\mu\alpha}\psi)+\\
&+2X\Theta W_{\alpha}\!=\!0\label{vi}\\
\nonumber
&\nabla_{\nu}\Theta\!-\!
2i(\overline{\psi}\boldsymbol{\sigma}_{\mu\nu}\boldsymbol{\pi}\boldsymbol{\nabla}^{\mu}\psi\!-\!
\boldsymbol{\nabla}^{\mu}\overline{\psi}\boldsymbol{\sigma}_{\mu\nu}\boldsymbol{\pi}\psi)-\\
&-2X\Phi W_{\nu}\!+\!2mS_{\nu}\!=\!0\label{ar}\\
\nonumber
&(\boldsymbol{\nabla}_{\alpha}\overline{\psi}\boldsymbol{\pi}\psi
\!-\!\overline{\psi}\boldsymbol{\pi}\boldsymbol{\nabla}_{\alpha}\psi)
\!-\!\frac{1}{2}\nabla^{\mu}M^{\rho\sigma}\varepsilon_{\rho\sigma\mu\alpha}+\\
&+2XW^{\mu}M_{\mu\alpha}\!=\!0\label{ai}
\end{eqnarray}
as Madelung-Gordon decompositions \cite{h2,k}: plugging the polar form (\ref{spinor}) or its derivative (\ref{decspinder}) gives expressions
\begin{eqnarray}
\nonumber
&-\nabla_{\mu}\ln{\phi}M^{\mu\sigma}
\!+\!\frac{1}{2}(\frac{1}{2}\nabla_{\mu}\beta\!-\!XW_{\mu})M_{\pi\nu}
\varepsilon^{\pi\nu\mu\sigma}-\\
\nonumber
&-(qA\!-\!\nabla\alpha)^{\sigma}\Phi
\!-\!\frac{1}{8}(\Omega\!-\!\partial\theta)^{\alpha\nu\rho}M_{\pi\kappa}
\varepsilon_{\alpha\nu\rho\mu}\varepsilon^{\pi\kappa\sigma\mu}+\\
&+\frac{1}{2}(\Omega\!-\!\partial\theta)_{\mu a}^{\phantom{\mu a}a}M^{\mu\sigma}\!-\!mU^{\sigma}\!=\!0\\
\nonumber
&-\nabla_{\sigma}\ln{\phi}\Phi\!+\!(\frac{1}{2}\nabla_{\sigma}\beta\!-\!XW_{\sigma})\Theta
\!+\!(qA\!-\!\nabla\alpha)^{\mu}M_{\mu\sigma}+\\
&+\frac{1}{4}(\Omega\!-\!\partial\theta)^{\alpha\nu\rho}\varepsilon_{\alpha\nu\rho\sigma}\Theta
\!+\!\frac{1}{2}(\Omega\!-\!\partial\theta)_{\sigma a}^{\phantom{\sigma a}a}\Phi\!=\!0\\
\nonumber
&\nabla_{\sigma}\ln{\phi}\Theta\!+\!(\frac{1}{2}\nabla_{\sigma}\beta\!-\!XW_{\sigma})\Phi-\\
\nonumber
&-\frac{1}{2}(qA\!-\!\nabla\alpha)^{\mu}M^{\pi\kappa}\varepsilon_{\pi\kappa\mu\sigma}
\!+\!\frac{1}{4}(\Omega\!-\!\partial\theta)^{\alpha\nu\rho}\varepsilon_{\alpha\nu\rho\sigma}\Phi-\\
&-\frac{1}{2}(\Omega\!-\!\partial\theta)_{\sigma a}^{\phantom{\sigma a}a}\Theta\!+\!mS_{\sigma}\!=\!0\\
\nonumber
&\frac{1}{2}\nabla_{\mu}\ln{\phi}M_{\pi\kappa}\varepsilon^{\pi\kappa\mu\sigma}
\!+\!(\frac{1}{2}\nabla_{\mu}\beta\!-\!XW_{\mu})M^{\mu\sigma}+\\
\nonumber
&+(qA\!-\!\nabla\alpha)^{\sigma}\Theta
\!+\!\frac{1}{4}(\Omega\!-\!\partial\theta)^{\alpha\nu\rho}M^{\mu\sigma}\varepsilon_{\alpha\nu\rho\mu}-\\
&-\frac{1}{4}(\Omega\!-\!\partial\theta)_{\mu a}^{\phantom{\mu a}a}M_{\pi\kappa}
\varepsilon^{\pi\kappa\mu\sigma}\!=\!0
\end{eqnarray}
as a straightforward substitution shows. The second and third, after plugging (\ref{M}), (\ref{S}, \ref{U}) and (\ref{T}, \ref{F}), become
\begin{eqnarray}
\nonumber
&\frac{1}{2}\nabla_{\alpha}\ln{\phi^{2}}\cos{\beta}
\!-\!(\frac{1}{2}\nabla_{\alpha}\beta\!-\!XW_{\alpha})\sin{\beta}+\\
\nonumber
&+(\nabla\alpha\!-\!qA)^{\mu}(u^{\rho}s^{\sigma}\varepsilon_{\rho\sigma\mu\alpha}\cos{\beta}
\!+\!u_{[\mu}s_{\alpha]}\sin{\beta})-\\
\nonumber
&-\frac{1}{2}(\Omega\!-\!\partial\theta)_{\alpha\mu}^{\phantom{\alpha\mu}\mu}\cos{\beta}-\\
&-\frac{1}{4}(\Omega\!-\!\partial\theta)^{\rho\sigma\mu}\varepsilon_{\rho\sigma\mu\alpha}\sin{\beta}\!=\!0\\
\nonumber
&\frac{1}{2}\nabla_{\nu}\ln{\phi^{2}}\sin{\beta}
\!+\!(\frac{1}{2}\nabla_{\nu}\beta\!-\!XW_{\nu})\cos{\beta}+\\
\nonumber
&+(\nabla\alpha\!-\!qA)^{\mu}(u^{\rho}s^{\sigma}\varepsilon_{\rho\sigma\mu\nu}\sin{\beta}\!-\!u_{[\mu}s_{\nu]}\cos{\beta})+\\
\nonumber
&+\frac{1}{4}(\Omega\!-\!\partial\theta)^{\rho\sigma\mu}\varepsilon_{\rho\sigma\mu\nu}\cos{\beta}-\\
&-\frac{1}{2}(\Omega\!-\!\partial\theta)_{\nu\mu}^{\phantom{\nu\mu}\mu}\sin{\beta}
\!+\!ms_{\nu}\!=\!0
\end{eqnarray}
and after diagonalization
\begin{eqnarray}
\nonumber
&\frac{1}{2}\varepsilon_{\mu\alpha\nu\iota}(\partial\theta\!-\!\Omega)^{\alpha\nu\iota}
\!-\!2(\nabla\alpha\!-\!qA)^{\iota}u_{[\iota}s_{\mu]}-\\
&-2XW_{\mu}\!+\!\nabla_{\mu}\beta\!+\!2s_{\mu}m\cos{\beta}\!=\!0\label{aux1}\\
\nonumber
&(\partial\theta\!-\!\Omega)_{\mu a}^{\phantom{\mu a}a}
\!-\!2(\nabla\alpha\!-\!qA)^{\rho}u^{\nu}s^{\alpha}\varepsilon_{\mu\rho\nu\alpha}+\\
&+2s_{\mu}m\sin{\beta}\!+\!\nabla_{\mu}\ln{\phi^{2}}\!=\!0\label{aux2}
\end{eqnarray}
in general. Next, take the left-hand side of Dirac spinor field equations: with (\ref{spinor}), using (\ref{aux1}, \ref{aux2}), and considering
\begin{eqnarray}
&i\boldsymbol{\gamma}^{\mu}\psi u^{\nu}s^{\alpha}\varepsilon_{\mu\rho\nu\alpha}
\!+\!u_{[\rho}s_{\mu]}\boldsymbol{\gamma}^{\mu}\boldsymbol{\pi}\psi
\!+\!\boldsymbol{\gamma}_{\rho}\psi=0\\
&is_{\mu}\boldsymbol{\gamma}^{\mu}\psi\sin{\beta}
\!+\!s_{\mu}\boldsymbol{\gamma}^{\mu}\boldsymbol{\pi}\psi\cos{\beta}\!+\!\psi=0
\end{eqnarray}
it is possible to see that
\begin{eqnarray}
\nonumber
&i\boldsymbol{\gamma}^{\mu}\boldsymbol{\nabla}_{\mu}\psi
\!-\!XW_{\sigma}\boldsymbol{\gamma}^{\sigma}\boldsymbol{\pi}\psi\!-\!m\psi=\\
\nonumber
&=-[i\boldsymbol{\gamma}^{\mu}(qA\!-\!\nabla\alpha)^{\rho}u^{\nu}s^{\alpha}\varepsilon_{\mu\rho\nu\alpha}+\\
\nonumber
&+(qA\!-\!\nabla\alpha)^{\iota}u_{[\iota}s_{\mu]}\boldsymbol{\gamma}^{\mu}\boldsymbol{\pi}
\!+\!(qA\!-\!\nabla\alpha)_{\mu}\boldsymbol{\gamma}^{\mu}+\\
&+is_{\mu}\boldsymbol{\gamma}^{\mu}m\sin{\beta}
\!+\!s_{\mu}\boldsymbol{\gamma}^{\mu}\boldsymbol{\pi}m\cos{\beta}\!+\!m\mathbb{I}]\psi\!=\!0
\end{eqnarray}
telling us that the Dirac spinor field equation is valid in general. Therefore (\ref{D}) is equivalent to the pair
\begin{eqnarray}
\nonumber
&\frac{1}{2}\varepsilon_{\mu\alpha\nu\iota}(\partial\theta\!-\!\Omega)^{\alpha\nu\iota}
\!-\!2(\nabla\alpha\!-\!qA)^{\iota}u_{[\iota}s_{\mu]}+\\
&+2(\nabla\beta/2\!-\!XW)_{\mu}\!+\!2s_{\mu}m\cos{\beta}\!=\!0\label{f1}\\
\nonumber
&(\partial\theta\!-\!\Omega)_{\mu a}^{\phantom{\mu a}a}
\!-\!2(\nabla\alpha\!-\!qA)^{\rho}u^{\nu}s^{\alpha}\varepsilon_{\mu\rho\nu\alpha}+\\
&+2s_{\mu}m\sin{\beta}\!+\!\nabla_{\mu}\ln{\phi^{2}}\!=\!0\label{f2}
\end{eqnarray}
as field equations for the module and the YT angle.

Further the geometric field equations are given by
\begin{eqnarray}
&\nabla_{\sigma}F^{\sigma\mu}\!=\!2q\phi^{2}u^{\mu}
\end{eqnarray}
alongside to
\begin{eqnarray}
\nonumber
&R^{\rho\sigma}\!-\!\frac{1}{2}Rg^{\rho\sigma}
\!=\!\frac{1}{2}[M^{2}(W^{\rho}W^{\sigma}\!\!-\!\!\frac{1}{2}W^{\alpha}W_{\alpha}g^{\rho\sigma})+\\
\nonumber
&+\frac{1}{4}(\partial W)^{2}g^{\rho\sigma}
\!-\!(\partial W)^{\sigma\alpha}(\partial W)^{\rho}_{\phantom{\rho}\alpha}+\\
\nonumber
&+\frac{1}{4}F^{2}g^{\rho\sigma}\!-\!F^{\rho\alpha}\!F^{\sigma}_{\phantom{\sigma}\alpha}-\\
\nonumber
&-\phi^{2}[(XW\!-\!\nabla\frac{\beta}{2})^{\sigma}s^{\rho}
\!+\!(XW\!-\!\nabla\frac{\beta}{2})^{\rho}s^{\sigma}+\\
\nonumber
&+(qA\!-\!\nabla\alpha)^{\sigma}u^{\rho}\!+\!(qA\!-\!\nabla\alpha)^{\rho}u^{\sigma}-\\
&-\frac{1}{4}(\Omega\!-\!\partial\theta)_{ij}^{\phantom{ij}\sigma}\varepsilon^{\rho ijk}s_{k}
\!-\!\frac{1}{4}(\Omega\!-\!\partial\theta)_{ij}^{\phantom{ij}\rho}\varepsilon^{\sigma ijk}s_{k}]]
\end{eqnarray}
with
\begin{eqnarray}
&\nabla_{\alpha}(\partial W)^{\alpha\mu}\!+\!M^{2}W^{\mu}\!=\!2X\phi^{2}s^{\mu}
\end{eqnarray}
so that the trace of the gravitational field equations and the divergence of the torsion field equations are
\begin{eqnarray}
&2R\!=\!M^{2}W^{2}\!-\!2m\phi^{2}\cos{\beta}
\end{eqnarray}
and
\begin{eqnarray}
&M^{2}\nabla W\!=\!4Xm\phi^{2}\sin{\beta}
\end{eqnarray}
which are obtained by employing the equations that we get after having $s^{\mu}$ dotted into (\ref{f1}) and (\ref{f2}) above.

This shows how module and YT angle link to curvature and torsion; at times this becomes a pure link between module and curvature and between YT angle and torsion.

In what follows we will proceed to deepen this link.
\section{Torsion Axial Vector and Yvon-Takabayashi angle}
In order to start a more concrete discussion, let us first reconsider (\ref{spinor}): in the rest frame and spin eigenstate, the spinor field is given according to the form
\begin{eqnarray}
&\!\psi\!=\!\phi e^{-i\alpha}e^{-\frac{i}{2}\beta\boldsymbol{\pi}}
\!\left(\!\begin{tabular}{c}
$1$\\
$0$\\
$1$\\
$0$
\end{tabular}\!\right)
\label{spinorspecial}
\end{eqnarray}
which can be used to infer a possible meaning of the YT angle in analogy with the phase. In QFT the $\alpha$ phase is related in terms of $\nabla_{\nu}\alpha\!=\!P_{\nu}$ to the momentum; similarly we could think that the YT angle can be written as
\begin{eqnarray}
&\nabla_{\nu}\beta/2\!=\!K_{\nu}
\end{eqnarray}
where $K_{\nu}$ could in total analogy be defined as the axial momentum of the travelling wave: intuitively, if we think at the momentum of the particle as the full momentum of the whole system, we may think at the axial momentum as the opposite momenta of the corresponding irreducible chiral components. In momentum space we may imagine this as an opposite translation along the third axis of the two point-like particles representing the two chiral parts, and in position space we may imagine this as an opposite rotation around the third axis of the two plane-fronted waves describing the two chiral projections. Then the YT angle is what maintains the two chiral parts independent.

From (\ref{spinorspecial}) we see that, for a spin eigenstate, a rotation around the third axis has the same effect of the unitary phase shift; so we may always assume $\alpha\!=\!0$ and write
\begin{eqnarray}
&\!\psi\!=\!\phi e^{-\frac{i}{2}\beta\boldsymbol{\pi}}
\!\left(\!\begin{tabular}{c}
$1$\\
$0$\\
$1$\\
$0$
\end{tabular}\!\right)
\label{spinorveryspecial}
\end{eqnarray}
in general. Neglecting the unitary phase is not a problem when considering single particle states: mathematically, if we consider the source of the gravitational field equations, that is the energy density tensor of the spinor field, it is possible to see that the time-time component receives identical contributions from phases and rotations around the third axis in the sense that we may write either
\begin{eqnarray}
&\partial_{t}\alpha\!=\!E
\end{eqnarray}
or 
\begin{eqnarray}
&-\frac{1}{2}\partial_{t}\theta^{12}\!=\!E
\end{eqnarray}
and get the same energy; also in the matter field equation nothing changes whether we consider the former or latter of the two above expressions \cite{Fabbri:2016laz}. This means that when the spinor is in the third axis spin eigenstate, information about the energy is encoded either in the phase or in the third axis rotation: both are equivalent and we may pick one at will. Here we will take the third axis rotation.

In absence of electrodynamics the Dirac equations are
\begin{eqnarray}
&\nabla_{\mu}\ln{\phi}\!+\!s_{\mu}m\sin{\beta}
\!-\!\frac{1}{2}\Omega_{\mu a}^{\phantom{\mu a}a}\!=\!0\\
&K_{\mu}\!+\!s_{\mu}(m\cos{\beta}\!-\!E)\!-\!XW_{\mu}
\!-\!\frac{1}{4}\varepsilon_{\mu\alpha\nu\iota}\Omega^{\alpha\nu\iota}\!=\!0
\end{eqnarray}
where we have used the definition of the axial momentum and the above constraint on energy. As a following step, we notice that if we were to neglect both torsion and the YT angle we would obtain that the above field equations would turn out to be reduced to the simplest form
\begin{eqnarray}
&\frac{1}{2}\Omega_{\mu a}^{\phantom{\mu a}a}\!=\!\nabla_{\mu}\ln{\phi}\label{g1}\\
&\frac{1}{4}\varepsilon_{\mu\alpha\nu\iota}\Omega^{\alpha\nu\iota}\!=\!s_{\mu}(m\!-\!E)\label{g2}
\end{eqnarray}
showing that the gravitational force acts in terms of two types of potentials: a first is the trace of the spin connection, or the divergence of the tetrads, and it gives rise to an effect of gravitational compression manifested on the gradient of the logarithm of the module, a second is the completely antisymmetric part of the spin connection, or the curl of the tetrads, and it gives rise to a gravitational tidal effect in the direction of the spin and proportional to the difference between mass and energy of the matter distributions \cite{Fabbri:2015xga}. Instead in the general situation where neither torsion nor the YT angle are neglected, defining the module $\phi\!=\!\phi_{0}e^{F}$ with $\phi_{0}$ solution of (\ref{g1}) and $F$ some function that vanishes when the YT angle vanishes, one can maintain the above effects of gravity while having
\begin{eqnarray}
&\nabla_{\mu}F\!=\!-s_{\mu}m\sin{\beta}\label{t1}\\
&K_{\mu}\!-\!ms_{\mu}(1\!-\!\cos{\beta})\!=\!XW_{\mu}\label{t2}
\end{eqnarray}
to describe the coupling between the torsion axial vector and the YT angle. Equation (\ref{t1}) gives $F$ in terms of the YT angle while (\ref{t2}) is the field equation describing how the torsion axial vector affects the YT angle dynamics.

The complete coupling is given when we also take
\begin{eqnarray}
&\nabla_{\alpha}(\partial W)^{\alpha\mu}\!+\!M^{2}W^{\mu}\!=\!2X\phi_{0}^{2}e^{2F}s^{\mu}\label{Q}
\end{eqnarray}
as the torsion field equations. They come with
\begin{eqnarray}
&M^{2}\nabla W\!=\!4Xm\phi_{0}^{2}e^{2F}\sin{\beta}\label{pcac}
\end{eqnarray}
as a partially-conserved axial-vector current in terms of which we see the explicit appearance of the YT angle.

Therefore the torsion axial vector is the potential of an interaction of the YT angle and, since this describes the relative motions of chiral components of the spinor field, then the torsion axial vector gives rise to a force between the chiral components of the spinorial field itself.

Because the YT angle appears as the argument of circular functions then we have to expect that some type of non-linearity will arise for the dynamics \cite{h4}.
\section{Zitterbewegung}
For a given spinor the chiral components are given by the left-handed and right-handed projections designated by the $L$ and $R$ spinors\footnote{This should not be confused with the Ricci scalar.}, and when the spinor is taken in the spin-up eigenstate it means that both $L$ and $R$ have the same spin; however, they have opposite chirality: so, they have opposite projections of the velocity. In taking care of seeing the dynamics of $L$ and $R$ then, we observe that these two spinors are not a stable system unless some factor intervenes for which the velocity of the irreducible chiral parts is inverted; maintaining the spin unchanged, this can be done by transmuting $L$ into $R$ and viceversa and in no other way. What this implies is that the torsion must switch $L$ and $R$ with one another, and therefore it must have the character of a scalar or pseudo-scalar field.

Although torsion is an axial vector, the fact that it is a massive field, thus subject to no gauge transformation, implies that the torsion pseudo-scalar component cannot be transformed away, and such a torsional pseudo-scalar part is exactly the one isolated by the partially-conserved axial-vector current: the torsion pseudo-scalar is thus the torsion axial vector divergence. As a matter of fact any pseudo-scalar or scalar would give similar dynamics, but the torsion need not be assumed as it is already present.

With torsion, and specifically its pseudo-scalar degree of freedom, giving an interaction between $L$ and $R$ parts, we should expect that $L$ and $R$ display some oscillatory behaviour, which could be seen, in the momentum space, as an oscillation along the third axis of two point-like particles, and in the position space, as an oscillation around the third axis of two plane-fronted waves in general.

To give a proper mathematical description, let us suppose that there is no relevant influence of gravity, so that expression (\ref{ds}) gives $\nabla s\!=\!0$ while from (\ref{g1}, \ref{g2}) we can deduce that the module $\phi_{0}$ is approximately constant and that $E\!=\!m$ is fixed: computing the divergence of the field equation (\ref{t2}) using the partially-conserved axial-vector current (\ref{pcac}) gives the following field equations
\begin{eqnarray}
&\nabla^{2}\beta\!+\!4m^{2}[(1\!-\!\cos{\beta})
\!-\!\frac{X}{m}W\!s\!-\!\frac{2X^{2}}{M^{2}}\frac{\phi^{2}}{m}]\sin{\beta}\!=\!0\label{sKG}
\end{eqnarray}
while dotting the field equation (\ref{Q}) into $s_{\mu}$ and employing (\ref{pcac}) gives the following field equations
\begin{eqnarray}
\nonumber
&\nabla^{2}W\!s\!-\!\frac{8X^{2}}{M^{2}}\phi^{2}m\cos{\beta}W\!s+\\
&+\frac{8X}{M^{2}}\phi^{2}m^{2}(\cos{\beta}\!-\!1)\!+\!M^{2}W\!s\!=\!-2X\phi^{2}
\end{eqnarray}
in general within our hypotheses. In the limit for $\beta$ small we have that $\phi$ is also constant and the two field equations above respectively reduce to the expressions
\begin{eqnarray}
&\nabla^{2}\beta
\!+\!4m^{2}(-\frac{X}{m}W\!s\!-\!\frac{2X^{2}}{M^{2}}\frac{\phi^{2}}{m})\beta\!=\!0\label{MH}
\end{eqnarray}
and
\begin{eqnarray}
&\nabla^{2}W\!s\!-\!\frac{8X^{2}}{M^{2}}\phi^{2}mW\!s\!+\!M^{2}W\!s\!=\!-2X\phi^{2}\label{L}
\end{eqnarray}
which are simpler and they can now be solved.

The simplest solution is given when $W\!s$ is also constant since in this case (\ref{L}) further reduces to 
\begin{eqnarray}
&-\frac{8X^{2}}{M^{2}}\phi^{2}mW\!s\!+\!M^{2}W\!s\!=\!-2X\phi^{2}\label{G}
\end{eqnarray}
which can be inverted as
\begin{eqnarray}
W\!s\!=\!\frac{2XM^{2}\phi^{2}}{8X^{2}\phi^{2}m-M^{4}}
\end{eqnarray}
furnishing the $W\!s$ constant in terms of the $\phi$ constant.

This can be plugged into (\ref{MH}) to finally yield
\begin{eqnarray}
\nabla^{2}\beta
\!+\!\left(\frac{8X^{2}\phi^{2}m}{M\sqrt{M^{4}-8X^{2}\phi^{2}m}}\right)^{2}\beta\!=\!0
\end{eqnarray}
showing that we do indeed obtain oscillatory behaviours for the YT angle whenever $M^{4}\!>\!8X^{2}\phi^{2}m$ is valid.

This stability condition $M^{4}\!>\!8X^{2}\phi^{2}m$ is certainly true for small values of the constant $m\phi^{2}$ and thus for any light enough particle or for small densities of the matter field.

The effective YT mass can be written as
\begin{eqnarray}
\mu\!=\!M\frac{8X^{2}\phi^{2}m/M^{4}}{\sqrt{1\!-\!8X^{2}\phi^{2}m/M^{4}}}
\end{eqnarray}
proportional to the torsion mass times a given correction.

Using the typical normalization $X^{2}\phi^{2}\!=\!mM^{2}$ gives
\begin{eqnarray}
\mu\!=\!M\frac{8m^{2}/M^{2}}{\sqrt{1\!-\!8m^{2}/M^{2}}}
\end{eqnarray}
with $M^{2}\!>\!8m^{2}$ as stability condition and the correction proportional to the ratio of the two masses of the model.

According to the standard wisdom, quantum field theory is known to have a cut-off beyond which new physics should enter into consideration, and what in the present analysis we seem to witness is that such a cut-off may be given by the torsion mass with the new physics entailed by the dynamical effects due to the YT angle.

On the other hand, (\ref{G}) is inverted also as
\begin{eqnarray}
2X^{2}\phi^{2}\!=\!\frac{M^{4}XW\!s}{4mXW\!s-M^{2}}
\end{eqnarray}
to make torsion explicit in terms of the module.

When plugged into (\ref{MH}) it eventually gives
\begin{eqnarray}
\nabla^{2}\beta
\!+\!\left(\frac{4mXW\!s}{\sqrt{M^{2}-4mXW\!s}}\right)^{2}\beta\!=\!0
\end{eqnarray}
yielding $M^{2}\!>\!4mXW\!s$ as alternative stability condition.

Stability condition $M^{2}\!>\!4mXW\!s$ is certainly true for torsion with large mass or small coupling to spinors but also when $XW\!s\!<\!0$ and therefore when the coupling of torsion to spinor fields is universally attractive.

In this case the effective YT mass is then
\begin{eqnarray}
\mu\!=\!\pm M\frac{4mXW\!s/M^{2}}{\sqrt{1-4mXW\!s/M^{2}}}
\end{eqnarray}
proportional to the torsion mass times a given correction.

We notice that for attractive coupling to spinors only the minus sign is allowed and in this case we have
\begin{eqnarray}
\mu\!=\!M\frac{4m|XW\!s|/M^{2}}{\sqrt{1+4m|XW\!s|/M^{2}}}
\end{eqnarray}
still valid even if the torsion mass is small or in the case in which its coupling to spinors is large.

In such case $\mu\!\approx\!2\sqrt{m}\sqrt{|XW\!s|}$ showing that the larger is the coupling the larger is the frequency of oscillation.

Notice that if the torsion-spin coupling is attractive no singular value is met, and therefore the expression for the effective Yvon-Takabayashi mass is always regular.

Of course, one should not limit oneself to this simplest solution, but discussing whether such a condition would hold in general would require solving the entire system of field equations, which is an extremely complicated procedure in general \cite{Fabbri:2016fxt}, and we leave it to future works.

We remark that the spinorial field (\ref{spinorveryspecial}) can equivalently be written in terms of the following expression
\begin{eqnarray}
&\!\psi\!=\!\phi 
\!\left(\!\begin{tabular}{c}
$e^{\frac{i}{2}\beta}$\\
$0$\\
$e^{-\frac{i}{2}\beta}$\\
$0$
\end{tabular}\!\right)
\end{eqnarray}
which in standard representation becomes
\begin{eqnarray}
&\!\psi\!=\!\phi \sqrt{2}
\!\left(\!\begin{tabular}{c}
$\cos{\frac{\beta}{2}}$\\
$0$\\
$-i\sin{\frac{\beta}{2}}$\\
$0$
\end{tabular}\!\right)
\end{eqnarray}
where the complex exponentials have been rearranged as circular functions. This expression is general, but taking small YT angles it reduces to the very specific form
\begin{eqnarray}
\!\psi\!\approx\!\frac{\phi}{\sqrt{2}}
\!\left(\!\begin{tabular}{c}
$1$\\
$0$\\
$-i\beta$\\
$0$
\end{tabular}\!\right)
\end{eqnarray}
showing that the so-called small component can literally be small only if the YT angle is itself small. These forms show that the small component, whether really small or not, is linked to the YT angle in general circumstances.

Then, the small component is also related to the effects of zitterbewegung. As a consequence, the YT angle itself is related to the presence of zitterbewegung effects.

And zitterbewegung phenomena should be expected in situations where $\beta$ has oscillatory behaviour \cite{Vaz:1993np}.
\section{Chirality}
The picture that has emerged is one for which a spinor, albeit fundamental, is not irreducible but instead can be decomposed in terms of $L$ and $R$ spinors, themselves irreducible, and such that they describe two matter distributions having opposite velocities, and whenever stability conditions are valid they display oscillatory motion; these stability conditions involve some requirement on the torsion dynamics, which is seen as the force acting between the two chiral components, and which induces their relative oscillation. In this way, torsion can be interpreted as the force that prevents the chiral components to separate.

This argument finds support in the fact that for massive torsion, and in the effective approximation 
\begin{eqnarray}
M^{2}W^{\mu}\!\approx\!2X\phi^{2}s^{\mu}
\end{eqnarray}
the Dirac field equations become
\begin{eqnarray}
&i\boldsymbol{\gamma}^{\mu}\boldsymbol{\nabla}_{\mu}\psi
\!+\!\frac{X^{2}}{M^{2}}\left(\overline{\psi}\psi\mathbb{I}
\!+\!i\overline{\psi}\boldsymbol{\pi}\psi
i\boldsymbol{\pi}\right)\psi\!-\!m\psi\!=\!0\label{NJL}
\end{eqnarray}
which is the Nambu--Jona-Lasinio model \cite{n--j-l}. This model is very well known, and it accounts for some prototypical version of the strong force in which the mechanism of the generation of mass of baryons was due to the breakdown of the symmetry under the chiral transformations.

As we said above, in terms of the Lounesto classification spinor fields can be classified in terms of the bi-linear quantities (\ref{1}, \ref{2}, \ref{3}, \ref{4}, \ref{5}) in six classes given by
\begin{eqnarray}
&\!\!\!\!\!\!\!\!1)\ \ \Phi\!\neq\!0\ \ \ \ \Theta\!\neq\!0\ \ \ \ 
M^{ab}\!\neq\!0\ \ \ \ S^{a}\!\neq\!0
\label{tipo1}\\
&\!\!\!\!\!\!\!\!2)\ \ \Phi\!\neq\!0\ \ \ \ \Theta\!=\!0\ \ \ \ 
M^{ab}\!\neq\!0\ \ \ \ S^{a}\!\neq\!0
\label{tipo2}\\
&\!\!\!\!\!\!\!\!3)\ \ \Phi\!=\!0\ \ \ \ \Theta\!\neq\!0\ \ \ \ 
M^{ab}\!\neq\!0\ \ \ \ S^{a}\!\neq\!0
\label{tipo3}\\
&\!\!\!\!\!\!\!\!4)\ \ \Phi\!=\!0\ \ \ \ \Theta\!=\!0\ \ \ \ 
M^{ab}\!\neq\!0\ \ \ \ S^{a}\!\neq\!0
\label{tipo4}\\
&\!\!\!\!\!\!\!\!5)\ \ \Phi\!=\!0\ \ \ \ \Theta\!=\!0\ \ \ \ 
M^{ab}\!\neq\!0\ \ \ \ S^{a}\!=\!0
\label{tipo5}\\
&\!\!\!\!\!\!\!\!6)\ \ \Phi\!=\!0\ \ \ \ \Theta\!=\!0\ \ \ \ 
M^{ab}\!=\!0\ \ \ \ S^{a}\!\neq\!0
\label{tipo6}
\end{eqnarray}
being clear that $U^a$ is always non-zero for a Dirac spinor field; notice that once the bi-linear quantities are known, the reconstruction theorem makes it possible to explicitly construct the spinor fields in each class \cite{T}. Heretofore, in the literature, except for the standard classical spinor fields, the types of regular spinor fields have been underestimated, for the relevance of singular spinor fields that encompass classes $4)$, $5)$, $6)$, in (\ref{tipo4}, \ref{tipo5}, \ref{tipo6}) \cite{daSilva:2012wp}.

Nevertheless, the splitting of regular spinor fields can be useful for a further physical analysis of equation (\ref{NJL}) and the equations that follow, an analysis that up to now has only resided in formal aspects in the literature.

Analogies with Yukawa potential and Higgs mechanism are clear when we consider the standard model after symmetry breaking \cite{Fabbri:2009ta}: neglecting vector boson interactions, the spinor field equations for the electron are
\begin{eqnarray}
&i\boldsymbol{\gamma}^{\mu}\boldsymbol{\nabla}_{\mu}\psi\!-\!YH\psi\!-\!m\psi\!=\!0
\end{eqnarray}
while the field equations for the Higgs are
\begin{eqnarray}
&\nabla^{2}H\!+\!N^{2}H\!+\!\frac{Y}{2}\overline{\psi}\psi\!=\!0
\label{Higgs}
\end{eqnarray}
with $N$ mass of the Higgs and $Y$ being the Yukawa coupling and having neglected higher-order Higgs terms.

In the effective approximation we have that
\begin{eqnarray}
&N^{2}H\!\approx\!-\frac{Y}{2}\overline{\psi}\psi
\label{effapprox}
\end{eqnarray}
so that the electron equations become 
\begin{eqnarray}
&i\boldsymbol{\gamma}^{\mu}\boldsymbol{\nabla}_{\mu}\psi
\!+\!\frac{1}{2}\frac{Y^{2}}{N^{2}}\overline{\psi}\psi\psi\!-\!m\psi\!=\!0
\label{effelectron}
\end{eqnarray}
showing in fact a strong analogy with (\ref{NJL}) above.

It is very important now to mention that equation (\ref{Higgs}) for type-3 regular spinor fields is led to the Klein-Gordon equation with no potential; the effective approximation given by (\ref{effapprox}) yields $N^{2}H\!\approx\!0$ and (\ref{effelectron}) yields the standard Dirac equation for type-3 spinor fields. This shows that the type-3 regular spinor fields would simply not admit a coupling to the Higgs field, mining its place in the Standard Model. But on the other hand, type-1 and type-2 regular spinor fields do, as in these cases the non-linear term in $\Phi Y^{2}/N^{2}$ does remain non-trivial. The Standard Model coupling to the Higgs field is an interaction that physically distinguishes between type-3 and other classes of regular spinor fields in Lounesto classification.

Because both torsion and the Higgs field have effective limit given by the NJL model, giving rise to an attraction between the chiral parts, then we may conclude that both torsion and the Higgs field can be seen as the mediators of the interaction in terms of which the chiral parts might form bound-states granting the stability of spinor fields.

Differences between torsion and Higgs field can be seen by re-writing their field equations in terms of the module and the YT angle: when this is done we end up with
\begin{eqnarray}
&i\boldsymbol{\gamma}^{\mu}\boldsymbol{\nabla}_{\mu}\psi\!+\!
\left(2\frac{X^{2}}{M^{2}}\phi^{2}e^{i\beta\boldsymbol{\pi}}
\!-\!m\mathbb{I}\right)\psi\!=\!0
\label{re-arr}
\end{eqnarray}
as well as
\begin{eqnarray}
&i\boldsymbol{\gamma}^{\mu}\boldsymbol{\nabla}_{\mu}\psi\!+\!
\left(\frac{Y^{2}}{N^{2}}\phi^{2}\cos{\beta}\!-\!m\right)\psi\!=\!0
\end{eqnarray}
showing that torsion would account for a richer dynamics in the chiral sector. Most notably, large oscillations of the angle might contemplate the value $\beta\!=\!\pi/2$ for which the Higgs-induced condensates would be vanished whereas a torsionally-induced condensate would still be present.

This is clear from the fact that (\ref{re-arr}) can be written as 
\begin{eqnarray}
&i\boldsymbol{\gamma}^{\mu}\boldsymbol{\nabla}_{\mu}\psi\!-\!
\left(2\frac{X^{2}}{M^{2}}\phi^{2}s^{\mu}\boldsymbol{\gamma}_{\mu}\boldsymbol{\pi}
\!+\!m\mathbb{I}\right)\psi\!=\!0
\end{eqnarray}
with no YT explicitly present and therefore such that for no value of $\beta$ we can get rid of the interaction term.

This last form also shows that an even richer dynamics can be possible if we do not take into account effective approximations, considering torsion as propagating.

For some general treatment about the dynamics of the torsionally-induced axial-vector currents we refer to \cite{daRocha:2007sd}.

As a consequence of this analysis, we have all elements to build an analogy between the above mentioned prototypical version of the strong force and our model in the effective approximation. No effective approximation might be considered and in this case the analogy would have to be built with the present-day description of strong interactions. The picture that emerges is one for which we may see any spinor as a chiral bound state due to torsion attractive mediation in the same way in which we see any baryon as a bound state for gluon attractive mediation.

This picture is quite simple and rather intuitive, which is the reason why we think it is appealing. But aside from the theoretical value, it may be used to infer one possible effect that could be of phenomenological importance.

In the following we will proceed to discuss it.
\section{Discussion}
Even though torsion is not a gauge field, the structural analogies between torsion and chromodynamics and the electroweak theory are interesting; but as it usually happens, more interesting still are their differences: the presence of a torsion mass in opposition to the masslessness of the gauge bosons is the most striking of all. One of its consequences is the existence of some partially-conserved axial-vector current for torsion in the form
\begin{eqnarray}
&M^{2}\nabla W\!=\!2X\Theta m
\end{eqnarray}
and this is what gives rise to the appearance of the above mentioned pseudo-scalar bound state; in it the divergence of torsion is the attractive potential which provides the well for the chiral binding. Massive chiral bound states are therefore possible even without radiative corrections.

Phenomenologically, the effects of this supplementary torsionally-induced attraction depend on the specific values of the torsion coupling constant, and even more on the torsion mass; however, a question we might already ask is whether there are systems of fermions where some additional attraction might play a role. To this question, a first answer that comes in mind is the recently increased tension existing between the theoretically calculated and experimentally established radius of the proton \cite{Pohl:2010zza}.

Because theoretical computations over-estimate the radius of the proton, we think it is quite natural to ask the question about whether such an extra binding force could actually give rise to corrections that bring the theoretical result back to be compatible with the observed value and consequently releasing the tension with experiments.

This question may have an answer that is much closer than what we think new physics could be because, despite torsion would indeed be new physics compared to normal paradigms, nonetheless torsion is already part of the most general geometric background of field theories.
\section{Conclusion}
In this paper, we have studied how the torsion massive axial vector field relates to the YT angle, showing in what way the divergence of the torsion massive axial vector has the role of an attractive force between the two chiral parts binding them into pseudo-scalar states; thus we may see a spinor as a chiral bound state due to torsion like we see a baryon as a bound state due to gluons: this analogy is justified by the fact that in the effective limit our model reduces to the Nambu--Jona-Lasinio model. A difference with the NJL model is that the torsion is already massive, and thus it has partially-conserved axial-vector currents, those giving rise to the pseudo-scalar bound state, as just mentioned; this is important because in our model spinor fields can be massive even without the quantum anomaly introduced by radiative processes. Some phenomenological effect of this extra binding force can be sought into a correction diminishing the theoretical value of the radius of the proton releasing the tension with measurements.

With no electrodynamics, approximating gravity away allowed us to find for the torsion massive axial vector a ground state in which the YT angle had oscillations when certain stability conditions were valid: this oscillation can be interpreted as the oscillation of the chiral parts about their equilibrium configuration. Such an oscillation is to be expected as a consequence of zitterbewegung effects.

Studying more general dynamics would involve leaving the assumption of small YT angle. Or allowing gravity.

And most in general allowing also electrodynamics.
\begin{acknowledgments}
Rold\~{a}o da Rocha is grateful to CNPq (Grant No. 303293/2015-2) and to FAPESP (Grant No.~2017/18897-8) for partial financial support.
\end{acknowledgments}


\begin{thebibliography}{40}
\bibitem{L}
P.Lounesto, \textit{Clifford Algebras and Spinors}\\
(Cambridge University Press, 2001).
\bibitem{Cavalcanti:2014wia}
R.T.Cavalcanti, ``Classification of Singular Spinor\\ Fields and Other Mass Dimension One Fermions'',\\ \textit{Int.J.Mod.Phys.D}\textbf{23}, 1444002 (2014).
\bibitem{Fabbri:2016msm}
L.Fabbri,
``A generally-relativistic gauge\\ classification of the Dirac fields'',\\
\textit{Int.J.Geom.Meth.Mod.Phys.}\textbf{13},1650078(2016).
\bibitem{j-l}
G.Jakobi, G.Lochak, 
``Introduction des parametres relativistes de Cayley-Klein dans la representation hydrodynamique de l'equation de Dirac'',\\
\textit{Comptes Rendus Acad.Sci.}\textbf{243}, 234, (1956).
\bibitem{h1}
D.Hestenes, ``Real Spinor Fields'',\\
\textit{J.Math.Phys.}\textbf{8}, 798 (1967).
\bibitem{Fabbri:2017pwp}
L.Fabbri, ``General Dynamics of Spinors'',
\textit{Adv. Appl. Clifford Algebras}\textbf{27}, 2901 (2017).
\bibitem{p-s}
M.E.Peskin, D.V.Schr\"{o}der, \textit{An Introduction to Quantum Field Theory} (Perseus Books, 1995).
\bibitem{Fabbri:2017fac}
L.Fabbri, ``Foundations Quadrilogy'', 1703.02287 [gr-qc].
\bibitem{daSilva:2012wp} 
J.M.Hoff da Silva, R.da Rocha,
``Unfolding Physics from the Algebraic Classification of Spinor Fields'',\\
\textit{Phys. Lett. B}\textbf{718}, 1519 (2013).
\bibitem{Ablamowicz:2014rpa}
R.Ab{\l}amowicz, I.Gon{\c c}alves, R.da Rocha,
``Bilinear Covariants and Spinor Fields Duality in Quantum Clifford Algebras'',
\textit{J. Math. Phys.}\textbf{55}, 103501 (2014).
\bibitem{daRocha:2016bil} 
R.da Rocha, R.T.Cavalcanti,
``Flag-dipole and flagpole spinor fluid flows in Kerr spacetimes'',\\
\textit{Phys. Atom. Nucl.} \textbf{80}, 329 (2017).
\bibitem{daRocha:2008we} 
R.da Rocha, J.M.Hoff da Silva,
``ELKO, flagpole and flag-dipole spinor fields, and the instanton Hopf fibration'',\\
\textit{Adv. Appl. Clifford Algebras} \textbf{20}, 847 (2010).
\bibitem{Villalobos:2015xca}
C.H.Coronado Villalobos, J.M.Hoff da Silva, R.da Rocha,\\
``Questing mass dimension 1 spinor fields'',\\
\textit{Eur.Phys.J.C} \textbf{75}, 266 (2015).
\bibitem{Cavalcanti:2014uta}
R.T.Cavalcanti,J.M.Hoff da Silva,R.da Rocha,``VSR symmetries in the DKP algebra: the interplay between Dirac and Elko spinor fields'',
\textit{Eur.Phys.J.Plus}\textbf{129}, 246 (2014).
\bibitem{daRocha:2013qhu}
R.da Rocha,L.Fabbri,J.M.Hoff da Silva,R.T.Cavalcanti, J.A.Silva-Neto,
``Flag-Dipole Spinor Fields in ESK Gravities'',
\textit{J.Math.Phys.}\textbf{54},102505(2013).
\bibitem{Fabbri:2014dxa}
L.Fabbri,``A discussion on the most general torsion-gravity with electrodynamics for Dirac spinor matter fields'',
\textit{Int.J.Geom.Meth.Mod.Phys.}\textbf{12},1550099(2015).
\bibitem{h2}
D.Hestenes, ``Local Observables in the Dirac Theory'',\\
\textit{J.Math.Phys.}\textbf{14}, 893 (1973).
\bibitem{k}
H.Krueger, ``Classical Limit of Real Dirac Theory:\\ Quantization of Relativistic Central Field Orbits'',\\
\textit{Found.Phys.}\textbf{23}, 1265 (1993).
\bibitem{Fabbri:2016laz}
L.Fabbri,
``Torsion Gravity for Dirac Fields'',\\
\textit{Int.J.Geom.Meth.Mod.Phys.}\textbf{14},1750037(2017).
\bibitem{Fabbri:2015xga}
L.Fabbri,
``Torsionally-gravitating charged matter fields and quanta'',
\textit{Gen.Rel.Grav.}\textbf{47}, 119 (2015).
\bibitem{h4}
D.Hestenes, ``Quantum mechanics from self-interaction'',\\
\textit{Found. Phys.}\textbf{15}, 63 (1985).
\bibitem{Fabbri:2016fxt}
L.Fabbri, ``Torsion Gravity for Dirac Particles'',\\
\textit{Int.J.Geom.Meth.Mod.Phys.}\textbf{14}, 1750127 (2017).
\bibitem{Vaz:1993np}
J.Vaz, W.A.Rodriguez, ``Zitterbewegung and the\\ electromagnetic field of the electron'',\\
\textit{Phys.Lett.B}\textbf{319}, 203 (1993).
\bibitem{n--j-l}
Y.Nambu, G.Jona-Lasinio, ``Dynamical model of elementary particles based on an analogy with superconductivity'', \textit{Phys.Rev.}\textbf{122}, 345 (1961).
\bibitem{T}
Y.Takahashi, ``Reconstruction of Spinor From Fierz Identities'', 
\textit{Phys.Rev.D}\textbf{26}, 2169 (1982).
\bibitem{Fabbri:2009ta}
L.Fabbri, ``The Spin-Torsion coupling and Causality for the Standard Model'',
\textit{Mod.Phys.Lett.A}\textbf{26}, 2091 (2011).
\bibitem{daRocha:2007sd}
R.da Rocha, J.G.Pereira, ``The Quadratic spinor\\ Lagrangian, axial torsion current, and\\ generalizations'',
\textit{Int.J.Mod.Phys.D}\textbf{16}, 1653 (2007).
\bibitem{Pohl:2010zza}
R.Pohl \textit{et al.}, ``The size of the proton'',\\
\textit{Nature} \textbf{466}, 213 (2010).
\end{thebibliography}
\end{document}